# People's Attitudes Toward Automated Vehicle and Transit Integration: Case Study of Small Urban Areas

Preprint[1]

Yu Song, Madhav V. Chitturi, Chris McCahill, David A. Noyce

University of Wisconsin-Madison

Please address correspondence to Yu Song: yu.song@wisc.edu

**Abstract**

Previous surveys of people's attitudes toward automated vehicles (AVs) and transit integration primarily took place in large urban areas. AV-transit integration also has a great potential in small urban areas. A survey of people's attitudes towards AV-transit integration was carried out in two small urban areas in Wisconsin, United States. A total of 266 finished responses were analyzed using text mining, factor analysis, and regression analysis. Results showed that respondents knew about AVs and driving assistance technologies. Respondents welcome AV-transit integration but were unsure about its potential impacts. Technology-savvy respondents were more positive but had more concerns about AV-transit integration than others. Respondents who enjoyed driving were not necessarily against transit, as they were more positive about AV-transit integration and were more willing to use automated buses than those who did not enjoy driving as much. Transit users were more positive toward AV-transit integration than non-transit users.

**Introduction**

Automated vehicles (AVs) are expected to become a disruptive force to public transit in the future. A RAND Corporation study stated that AVs have an advantage in operating costs compared with mass transit and paratransit, and that substituting those services with AVs will improve social welfare (Anderson et al. 2014). A model-based quantitative study by Levin and Boyles predicted that demand for transit will decrease when more people have access to AVs (Levin and Boyles 2015). Some researchers hold the opinion that while AVs may eliminate the need for conventional transit services, high capacity transit will still be needed on major travel corridors, with AVs serving as a support (Martinez and Crist 2015; TRB 2017; Currie 2018). The increasing private ownership of AVs and infrastructure may change the situation of subsidy and investment for public transit (Grush and Niles 2016).

Transit could take various forms in a more automated world. Transit may be automated itself or integrated with AVs that provide feeder or "first-mile and last-mile" services. New operational strategies and technologies may be applied to improve service efficiency and quality. As with any type of new technology, public attitudes and opinions are an important factor for the planning, designing, implementation and more importantly adoption of an AV-transit integrated system. Public attitudes and opinions determine the broader adoption of such an integration.

The purpose of this study is to find out what the general public in small urban areas think about AVs and transit, and what are the relationships between their demographic features, travel habits, attitudes, and their opinions. Small urban areas are defined as urban areas with a population below 200,000 (OECD 2013). This study was carried out in two small urban areas in Wisconsin, the Eau Claire area (2010 population: 161,151) and the Janesville area (2010 population: 160,092). Online questionnaire surveys were distributed, asking questions about people's travel habits, opinions about AVs, and their attitudes toward technology, driving, and transit.

Two facts motivated this study. First, while there are many previous studies about people's opinions toward AVs that were carried out in large urban areas, few studies have considered small urban areas, which have very different travel patterns and transit services from large urban areas. Second, the small urban areas selected for this study share a lot of transportation characteristics with many other small urban areas in the United States. About 64% of the urban areas in the United States are small urban areas, making them a significant and essential part of the future AV market.

**Literature Review**

It is unclear how automation will change how transit operates and what form of vehicles transit services will take. A framework of a future transportation system with publicly shared AVs and personal AVs was proposed (Liu, Fagnant, and Zhang 2016; Liu 2016). Based on this framework, there are three forms of AV-transit integration.
- Shared AV systems, which are similar to a taxi system, ride-hailing systems of transportation network companies, or on-demand microtransit systems (Fagnant and Kockelman 2014; 2018; Azevedo et al. 2016; Chen and Kockelman 2016; Farhan and Chen 2018).
- AVs as first-mile and last-mile services, which feed and pick up passengers to and from transit stations along corridors served by high-capacity transit such as bus rapid transit and urban rail transit (Chong et al. 2011; Moorthy et al. 2017; Shen, Zhang, and Zhao 2018; Scheltes and de Almeida Correia 2017).
- Automated high-capacity transit systems (Alessandrini et al. 2015; Lazarus et al. 2018).

Previous studies, published within the most recent five years, about the public's attitude and potential consumers' preferences toward AVs, covered privately-owned AVs and their technologies (Payre, Cestac, and Delhomme 2014; Schoettle and Sivak 2014a; 2014b; Daziano, Sarrias, and Leard 2017; Bansal and Kockelman 2017; Bansal, Kockelman, and Singh 2016; Haboucha, Ishaq, and Shiftan 2017), and multiple types of AV-transit integration (Bansal, Kockelman, and Singh 2016; Haboucha, Ishaq, and Shiftan 2017; Lavieri et al. 2017; Krueger, Rashidi, and Rose 2016; Lu et al. 2017; Yap, Correia, and van Arem 2016; Dong, DiScenna, and Guerra 2019).

These studies generally used surveys to collect public opinions toward AV technologies. Overall, there are more people holding positive opinions toward vehicle automation technologies, but the general public still have concerns about safety when riding an AV, the inability to manually control AVs, system failure, and data privacy (Schoettle and Sivak 2014a; 2014b). Preferences and attitudes toward AVs vary among different groups of people and different geographical locations (Payre, Cestac, and Delhomme 2014; Schoettle and Sivak 2014a; 2014b; Haboucha, Ishaq, and Shiftan 2017). Two surveys, one done in New York and the other in Austin, Texas, showed that the amount of money people willing to pay for vehicle automation technologies ranged from $4,000 to $7,000, depending on the level of automation (Daziano, Sarrias, and Leard 2017; Bansal, Kockelman, and Singh 2016).

Regarding coverage on AV-transit integration, most of previous studies evaluated public opinions toward shared AV systems (Bansal, Kockelman, and Singh 2016; Haboucha, Ishaq, and Shiftan 2017; Lavieri et al. 2017; Krueger, Rashidi, and Rose 2016), two studies covered AV as first and last mile services (Lu et al. 2017; Yap, Correia, and van Arem 2016), and one study covered automated buses (Dong, DiScenna, and Guerra 2019).

- For a shared AV system, cost is a major factor affecting personal use (Bansal, Kockelman, and Singh 2016). Young and tech-savvy people are more likely to use a shared AV system (Haboucha, Ishaq, and Shiftan 2017; Lavieri et al. 2017; Krueger, Rashidi, and Rose 2016). Frequent transit users are no more likely to use a shared AV system than frequent car users (Lavieri et al. 2017).
- People do not fully trust vehicle automation technologies, but they generally would like to see AVs providing first-mile and last-mile services (Lu et al. 2017; Yap, Correia, and van Arem 2016).
- Regarding automated buses, a survey study showed that only 13% of the respondents would agree to ride a bus without a human operator onboard (Dong, DiScenna, and Guerra 2019). Male respondents tend to be more likely to ride a driverless bus than female respondents, and younger respondents (18-34 years' old) are more willing to ride driverless buses than respondents in older age groups.

In summary, previous studies found that people are generally willing to use AV services, either as a replacement or a complement of transit services, but their preferences for different technologies and types of service vary. Time and money costs are often critical factors affecting people's willingness to use AV-integrated transit services.

**Data**

A questionnaire was designed using Qualtrics and published online. Responses were collected in the Eau Claire area between April 16 and June 10 and in the Janesville area between May 1 and June 30, 2019. Both city governments helped distribute the survey. A total of 266 complete responses were obtained and used for statistical analysis, including 216 from Eau Claire and 50 from Janesville. As the two cities were very similar in terms of population and transit services, we combined survey responses from the two cities for analysis.

The online questionnaire was designed to understand how well the existing transit systems serve communities in the two small urban areas, the role that emerging technologies like automated driving could play, and how to better plan for the future. A total of 30 questions were asked, which covered the following aspects.

- Exposure and opinions about vehicle automation and driving assistance technologies.
- Transit usage.
- Travel habits.
- Attitudes toward technologies, driving, and transit.
- Demographic information.

An open-ended question was included in the survey to ask about people's general opinion about future of AV-transit integration. Out of the 266 finished responses, 200 included a written comment in response to this open-ended question. These comments were analyzed using a text mining method and quantified based on the emotion reflected.

Respondents rated their level of agreement with three sets of statements to gauge their general attitudes toward technology, driving, and transit. Each set contained four to six attitudinal statements, shown in Figure 1. These statements were adapted and modified from Haboucha, Ishaq, and Shiftan (2017). A five-level Likert scale was used to measure the level of agreement, from "strongly agree," "agree," "neither agree nor disagree," "disagree," to "strongly disagree." Responses were encoded as an integer between -2 and 2 (i.e., -2, -1, 0, 1, or 2), in which a larger number corresponds with a more positive attitude. For example, the statement, "I try new products before my friends and neighbors," is positively correlated with the attitude toward technology, thus a "strongly agree" rating was assigned a "2" while a "strongly disagree" rating was assigned a "-2." The statement, "I have little to no interest in new technology," is negatively correlated with the attitude toward technology, thus a "strong agree" rating was assigned a "-2" while a "strongly disagree" rating was assigned a "2".

Aside from the open-ended question and the attitudinal questions, other questions were designed as multiple choices. Those responses were encoded as ordinal variables or 0-1 indicators, based on the design of specific questions and options. These variables are summarized in the "Survey Results" section. Descriptive statistics of all variables used in the regression analysis were provided in the "Statistical Analysis" section.

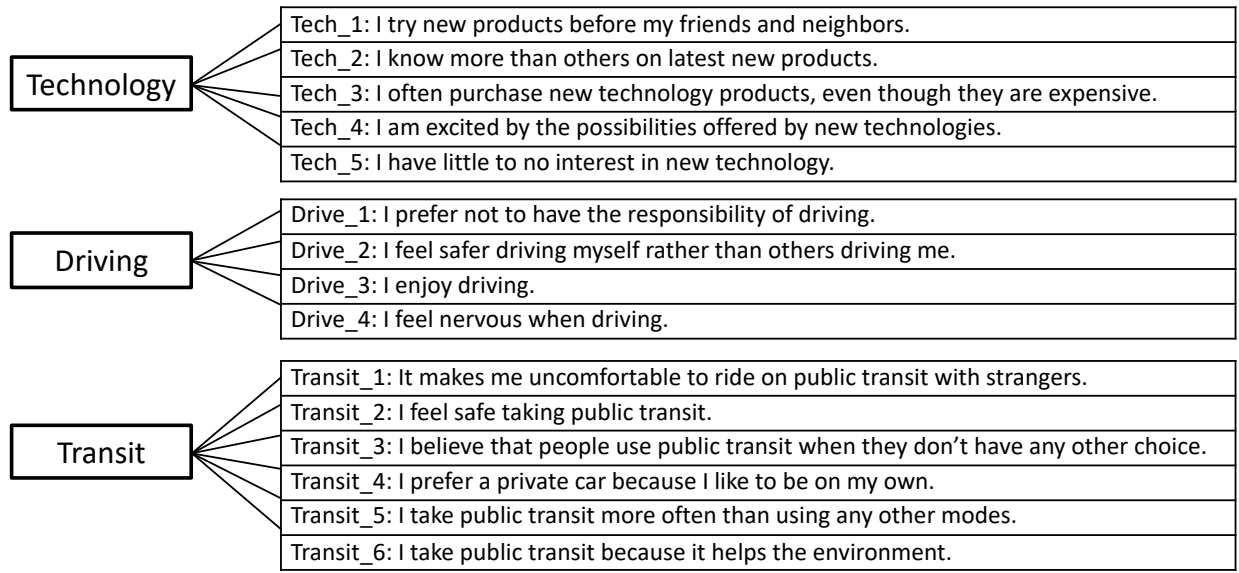

Figure 1 Attitudinal statements

## Methodology

We first aggregated and summarized the survey results to provide an overview of the respondents' demographics, travel pattern, and opinions. We then used text mining, confirmatory analysis, and regression analysis for an in-depth understanding of people's attitudes toward AV-transit integration. A text mining technique, sentimental analysis, was used to quantify and analyze text comments. A confirmatory factor analysis was conducted to quantify and evaluate analyze the attitudinal responses. A regression analysis was carried out to evaluate the relationship between people's opinions and factors including demographic features, travel patterns, and attitudes toward technology, driving, and transit.

*Text Mining*

Respondents provided written opinions regarding the future of AVs and transit. Responses ranged from 1 word to 118 words, with an average of 23 words (standard deviation: 22.4). A sentiment analysis (or "opinion mining") was used to apply a quantitative measure to each comment indicating how positive or negative it could be characterized as.

Packages 'tidytext', 'tidyr', and 'dplyr' were used in R to conduct the sentiment analysis. The comments were tidied by removing stop words, punctuations, and numbers. Then, the tidied comments were tokenized, which means each word in a tidied comment was pulled out and put on a single row, with its comment ID marked in an adjacent column. A sentiment score was then developed for each comment by matching the words in the comment to a developed lexicon. There are multiple lexicons available for sentiment analysis, most of which use a "-1, 0, 1" scale to quantify the emotion associated with a word, indicating negative, neutral, and positive sentiments. The AFINN lexicon was used for this study because this lexicon is one of the simplest and most popular lexicons for sentiment analysis (Nielsen 2011). AFINN lexicon (Version: AFINN-en-165.txt) included 3,382 manually labeled English words, each assigned with a sentiment score ranging from -5 to 5 (Silge and Robinson 2017). The sentiment score for a comment is the sum of the scores assigned to the matched words in that comment. For this study's analysis, two scales of sentiment scores were used. The first one is a normalized sentiment score, calculated by dividing the sum sentiment score for each comment by the comment's tidied wordcount (wordcount after the tidying process). The second one is a simplified, "-1, 0, 1"-scale score. The 66 responses (out of 266) with no response to the open-ended question were assigned a sentiment score of 0, assuming that "no comment" means neutral sentiment.

*Confirmatory Factor Analysis*

The attitudinal questions are meant to capture respondents' perspectives in three distinct areas. Five questions reflect the respondents' attitudes toward technology, four reflect the attitudes toward driving, and six reflect the attitudes toward transit. In factor analysis, each response is considered a "direct measure" of agreement with a particular statement. Collectively, these statements reflect respondents' general attitudes in each area, which are called "underlying factors" or "latent factors" and cannot be measured directly. For example, respondents can answer a series of questions related to their attitudes toward technology better than they can quantify their own general tech-savviness.

A confirmatory factor analysis (CFA) could help estimate the relationship between underlying factors and direct measures, as well as factor scores (Brown 2015). In CFA, the relationship between underlying factors and direct measures is defined as:

$$X = \beta F + \epsilon \qquad (1)$$

where, $X$ is a vector of direct measures; $F$ is a vector of underlying factors; $\beta$ is the loadings of underlying factors to the measures; and $\epsilon$ is the error term.

The lavaan package in R was used to carry out the CFA. The CFA used a maximum likelihood (ML) estimation procedure to estimate the factor loadings. A factor loading is analogous to a correlation factor, which shows the variance explained by the directly measured variable (i.e., respondents' level of agreement to a statement) on the underlying factor (i.e., attitude). For each response, a factor score was calculated as a weighted average:

$$s_i = \frac{\sum_1^n (\beta_{i,n} r_{i,n})}{n} \qquad (2)$$

where, $s_i$ is a response's score for factor (i.e., attitudinal aspect) $i$, with $i$ = technology, driving, or transit; $\beta_{i,n}$ is the estimated factor loading of attitudinal aspect $i$ on the level of agreement to statement $n$ in group $i$, the sign of a factor loading is pre-determined by the design of the statement to reflect its positive and negative relationship with the corresponding attitudinal aspect; and $r_{i,n}$ is the respondent's level of agreement to statement $n$.

*Regression Analysis*

Linear regression and ordered logistic regression were used to evaluate the effects of people's attitudes toward technology, driving, and transit, on their opinions about AVs and transit. Three regression models were developed for three response variables: normalized sentiment ("normalized"), simplified sentiment ("simplified"), and the willingness to ride automated transit vehicle without an operator onboard ("av_bus_nd"). We treated "normalized" as a continuous variable and used linear regression for the modeling. The other two response variables "simplified" and "av_bus_nd", were modeled as ordinal variables, using ordered logistic regression.

Explanatory variables included in the regression analysis captured the respondents' exposure to existing driving assistance technology, their travel habits, demographics, as well as their attitudes toward technology, driving, and transit. The same set of explanatory variables were used for all the four models.

**Survey Responses**

The survey results indicate that the respondents are well-exposed to information about AVs and driving assistance technologies. They generally welcome vehicle automation and driving assistance technologies but are still unsure about the future of vehicles equipped with more of these technologies. Most of the respondents have access to private vehicles and are satisfied with driving them. Transit was not one of the most popular daily traveling modes for these respondents, with the common reasons being "not convenient" and "not flexible" for their needs. While most respondents felt transit was generally safe, they also felt transit is primarily for people without access to other transportation modes. People are uncertain about the safety benefits of adding automation technologies to transit. At the current stage, they still prefer having a human operator even if transit vehicles were automated.

*Demographics*

Respondents are well distributed across all age groups, with more respondents (25%) from the 25-34 age group than from the others. There are more female (66%) respondents than from other gender groups. Most of the respondents (95%) have a valid driver license. Most of the respondents (94%) received some college education or higher. The average household size is 2.4 people, and 75% of the respondents do not have young kids (< 12 years) in their household.

*Exposure and Opinions about Vehicle Technologies*

In terms of vehicle automation and driving assistance technologies, 94% of the respondents have heard of AVs, and 91% of the respondents have a vehicle with one or more driving assistance technologies. The most common technologies the respondents have on their vehicles are cruise control, blind spot detection and warning, and lane departure avoidance. In general, the respondents are happy about those driving assistance technologies. About three quarters of the respondents would like more driving assistance technologies on their next car purchase.

*Transit Usage*

More than three quarters of the respondents never use transit. The average number of transit trips in a week, was less than 1 (0.91). The top three concerns in using transit for the respondents were lack of convenience, lack of flexibility, and poor access. While a considerable proportion of the respondents (47%) are unsure about whether driving assistance technologies will improve the safety for transit vehicles, 34% are positive and 17% are negative about the safety benefits of driving assistance technologies. Over three quarters of the respondents have concerns about taking a fully automated transit vehicle with no human operator onboard, with 48% of the respondents not feeling comfortable riding such a vehicle, and 27% of the respondents unsure about it. However, for the case of a mostly automated transit vehicle with a human operator onboard, about 70% of the respondents would feel comfortable taking it.

### *Travel Habits*

Travel habit questions reflect household vehicle ownership, commute pattern, and mode choice. Over 94% of the respondents have one or more personal vehicles in their household. Over 83% of the respondents commute to work, with 73% of the respondents commuting five times or more per week. Half of the respondents have a commute shorter than 5 miles and 67% of the respondents have a one-way commute time of less than 20 minutes. Driving a personal vehicle was the first mode choice for 68% of the respondents' commute trips and 86% for other trips. Transit was the first choice for commute trips to a little more than 7% of the respondents and less than 4% of the respondents for other trips.

### *Attitudes toward Technology, Driving, and Transit*

As describe earlier, attitudinal questions cover three aspects: technology, driving, and transit. The responses imply that the respondents are generally both interested and excited in new technologies (about 70% agree or strongly agree), but they are somewhat more reserved in their willingness to spend money on such new technologies. The responses to questions about driving indicated that the respondents are satisfied with driving for their daily travels, and they feel confident and comfortable when driving vehicles themselves. About 60% of respondents enjoy driving and feel safer driving themselves than when driven by others. Over 50% of the respondents implied a preference for using a personal vehicle to taking transit. Respondents generally feel comfortable and safe when taking transit (about 60% agree or strongly agree), but about half of them believe that people only take transit due to lack of access to other modes of transportation.

## **Statistical Analysis Results**

The analysis consists of three parts, a sentimental analysis of text comments to summarize people's opinions, a CFA of Likert-scale responses to quantify attitudes, and a regression analysis to understand the relationships among all variables. Analysis results are as follows.

### *Sentiments*

There were 39 (15%) comments with negative sentiment, 166 (62%) comments with neutral (zero) sentiment, and 61 (23%) comments with positive sentiment. Based on the positive and negative components of sentiment scores, the responses can be categorized into three groups and six subgroups, as listed in Table 1. This distribution shows that most of the respondents (90%) were certain about their opinions, using words with only negative, neutral, or positive sentiment in their comments. The rest respondents (10%) used a mixture of words with negative and positive sentiments.

Table 1  Sentiment groups

| Group (number, %) | Subgroup (number, %) | Definition |
| --- | --- | --- |
| Negative (39, 15%) | Negative only (28, 11%) | Negative sentiment consisting of only a negative score |
|  | Mainly negative (11, 4%) | Negative sentiment consisting of a negative and a positive score |
| Neutral (166, 62%) | Zero (160, 60%) | Zero sentiment consisting of no negative of positive score |
|  | Canceled out (6, 2%) | Zero sentiment consisting of a negative and a positive score |
| Positive (61, 23%) | Positive only (51, 19%) | Positive sentiment consisting of only a positive score |
|  | Mainly positive (10, 4%) | Positive sentiment consisting of a negative and a positive score |
| **Total (266, 100%)** | | |

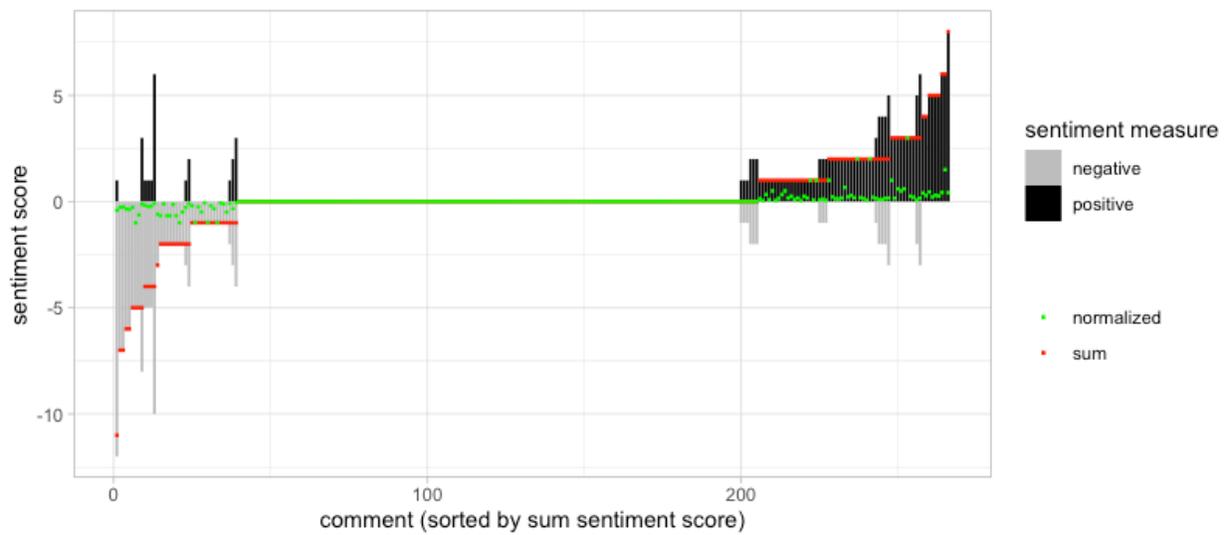

a. Sentiment scores by comment

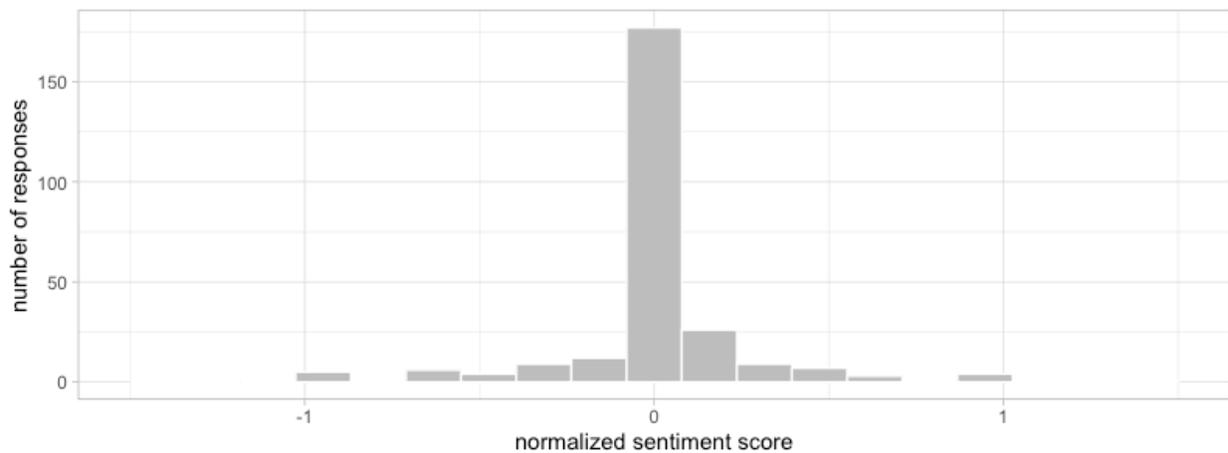

b. Normalized sentiment score distribution

**Figure 2  Sentiment scores**

Figure 2a visualizes the positive and negative components of sentiment score by comment, with the corresponding summed sentiment scores and normalized sentiment scores for comparison. Figure 2b illustrates the distribution of normalized sentiment scores. The summed sentiment score ("sum") ranges from -11 to 8, with an average of 0.11 and a standard deviation of 1.92. The normalized sentiment score ("normalized") ranges from -1 to 3, with an average of 0.01 and a standard deviation of 0.24. The distributions indicate that respondents' opinion about the future of AVs and transit tends to be slightly positive. However, the fact that 64% of the respondents' sentiment scores are zero means that people are still very unsure about how they think of vehicle automation and its potential effect on future transit.

The words in comments were grouped according to their sentimental values, and were ranked by their contribution (i.e., occurrences). Figure 3 illustrated the ranking of words in each sentiment value group. By looking at the words with high contribution to their sentimental value groups, a pattern is found that safety ("accidents", "safety", and "safe") is raising concerns about AV-transit integration, while comfort ("comfortable") and potential opportunities ("opportunity") are attractive features that may be winning respondents' interest about AV-transit integration.

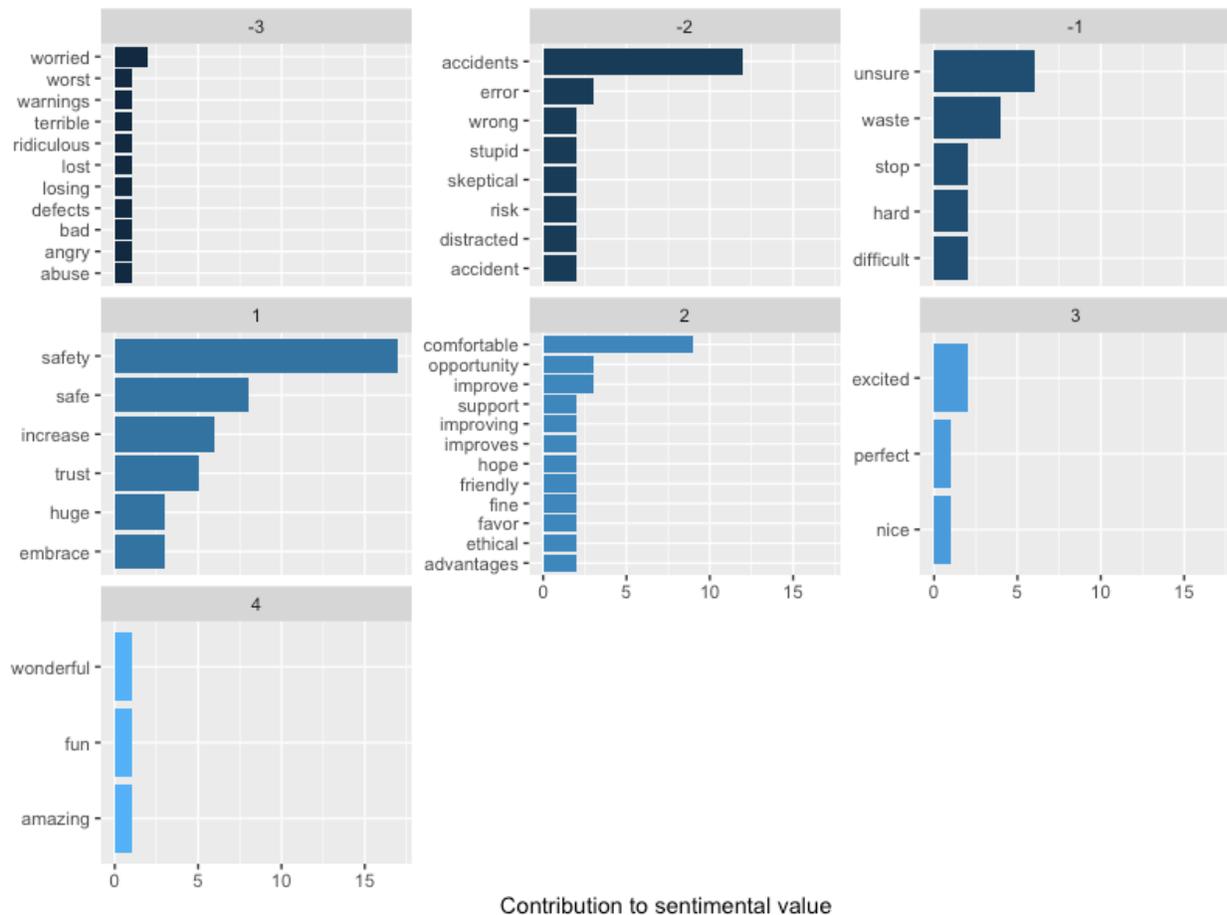

**Figure 3  Words in comments ranked by contribution for each sentiment value**

*Attitudes*

CFA was used to quantify respondents' attitudes toward technology, driving, and transit. Before carrying out CFA directly with all the collected attitudinal data, it is helpful to run an exploratory factor analysis (EFA) to verify the designed construct of the underlying attitudinal factors and direct measures (Swisher, Beckstead, and Bebeau 2004). Selection of direct measures to use for CFA can be adjusted with the information from EFA. An EFA was carried out with the 226 encoded responses to the 15 attitudinal questions. The EFA suggested that five factors were the least needed to fit the model sufficiently well. We expected that three major factors of the five EFA factors would match well with the three underlying factors (technology, driving, and transit) that we designed for the survey. Two other factors would be factors that could not be captured by the construct design.

Factor loadings from the EFA are listed in Table 2. A criterion we used to determine if a measure is primarily loaded to a factor is that the measure's absolute factor loading is the largest across the row. By comparing the factor loadings in Table 2, we found that most of the "technology" attitudinal measures were loaded to Factor 1, all "driving" attitudinal measures were loaded to Factor 2, and most of the "transit" attitudinal measures were loaded to Factor 3. This observation supports an inference that Factors 1, 2, and 3 are factors "technology", "driving", and "transit", respectively. Factors 4 and 5 are unknown factors that could not be captured by the designed construct. The attitudinal statements, Tech_4: "I am excited by the possibilities brought by new technologies" and Tech_5: "I have little to no interest in new technology", are not as closely related to a behavior in purchasing and using new technology products as the other three statements in the same set. Similarly, Transit_3: "I take public transit more often than using any other modes" is more about transit usage, and Transit_4: "I take public transit because it helps the environment" is more about environment-related attitude rather than attitude toward transit. Based on the EFA results, the measures that were primarily loaded to Factors 1, 2, and 3 were used to fit the CFA model.

**Table 2 Exploratory factor analysis results**

| Factor | Measure | Factor1 | Factor2 | Factor3 | Factor4 | Factor5 |
|---|---|---|---|---|---|---|
| Technology | Tech_1 | **0.870** | | | 0.139 | |
| | Tech_2 | **0.730** | 0.105 | | 0.257 | |
| | Tech_3 | **0.687** | | -0.110 | 0.160 | |
| | Tech_4 | 0.340 | | | **0.788** | |
| | Tech_5 | 0.287 | | | **0.784** | -0.115 |
| Driving | Drive_1 | | **0.577** | -0.115 | | -0.174 |
| | Drive_2 | 0.140 | **0.412** | -0.247 | -0.214 | -0.164 |
| | Drive_3 | | **0.942** | | 0.100 | 0.140 |
| | Drive_4 | 0.135 | **0.572** | | | -0.279 |
| Transit | Transit_1 | | | **0.891** | | |
| | Transit_2 | | | **0.535** | | 0.219 |
| | Transit_3 | | | **0.249** | | |
| | Transit_4 | | -0.340 | **0.545** | | 0.135 |
| | Transit_5 | 0.101 | -0.185 | 0.183 | | **0.705** |
| | Transit_6 | | -0.139 | 0.325 | | **0.525** |

Note: Test of the hypothesis that 5 factors are sufficient.
    Chi-squared: 35.81 on 40 degrees of freedom; p-value = 0.659.
    Boldfaced factor loadings have the largest absolute values in their rows.

The CFA model is shown in Table 3. Based on the model fit indices, the CFA model fits moderately well to the data. The comparative fit index is 0.946 (> 0.90), the Tucker-Lewis index (TLI) is 0.928 (> 0.90 but < 0.95), the root mean square error of approximation (RMSEA) is 0.060 (< 0.08), and the standardized root mean square residual (SRMR) is 0.062 (<0.08). All factor loading estimates were statistically significant at the 0.05 level. Each factor loading was standardized to a value between 0 and 1.

Factor scores were calculated using the estimated factor loadings as weights. The distributions of factor scores for the three attitudinal aspects are shown in Figure 4. "Technology" has an average score of -0.12 (s.d. = 0.76, min = -1.66, max = 1.66). "Driving" has an average score of 0.06 (s.d. = 0.30, min = -1.26, max = 1.26). "Transit" has an average score of 0.06 (s.d. = 0.35, min = -0.78, max = 0.96).

**Table 3  Confirmatory factor analysis results**

| Factor | Measure | Loading Estimate | Standard Error |
|---|---|---|---|
| Technology | Tech_1 | 0.917 | 0.058 |
|  | Tech_2 | 0.761 | 0.057 |
|  | Tech_3 | 0.816 | 0.067 |
| Driving | Drive_1 | 0.797 | 0.081 |
|  | Drive_2 | 0.490 | 0.070 |
|  | Drive_3 | 0.896 | 0.067 |
|  | Drive_4 | 0.662 | 0.070 |
| Transit | Transit_1 | 0.822 | 0.077 |
|  | Transit_2 | 0.497 | 0.059 |
|  | Transit_3 | 0.258 | 0.071 |
|  | Transit_4 | 0.718 | 0.074 |
| **Model fit indices:** CFI: 0.946; TLI: 0.928; RMSEA: 0.060; SRMR: 0.062 | | | |

Note: All estimates are statistically significant at the 0.05 level.
CFI = confirmatory factor index; TLI = Tucker Lewis index;
RMSEA = root mean square error of approximation; and
SRMR = standardized root mean square residual.

*Regressions*

Regression analysis was carried out to obtain insights regarding two variables of interest from the survey, people's opinions about AV-transit integration and people's willingness to take an automated bus without a human operator onboard. Two scales of measure were set up for people's opinion, normalized sentiment ("normalized"), and simplified sentiment ("simplified"). People's response regarding automated bus was measured by the variable "av_bus_nd". Descriptive statistics of variables used in the analysis are summarized in Table 4.

Linear regression was used to model the normalized sentiment score ("normalized") but a good model fit could not be achieved. The linear regression models are summarized in Table 5. Although the models did not fit well, there are a few statistically significant variables (at 0.2 level or lower), which is worth attention. Respondents' attitudes toward technology and driving (the "tech" and "driving" variables) are positively and negatively correlated with "normalized," respectively, indicating that tech-savviness may be leading to respondents' more concerns about AV-transit integration, but people who love driving may be more positive about AV-transit integration than those not as interested in driving. The "datech_happy" variable is positively correlated with

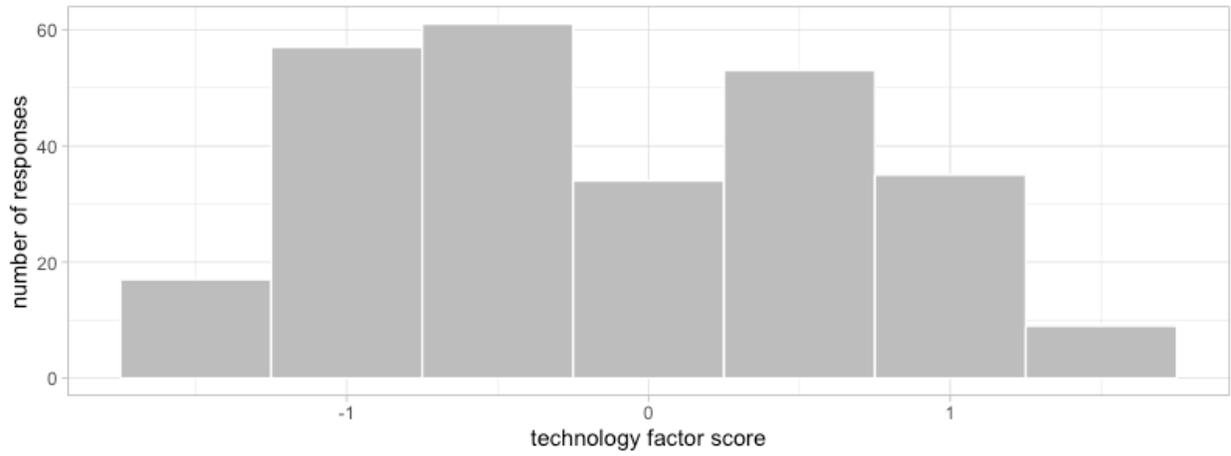
a. Technology

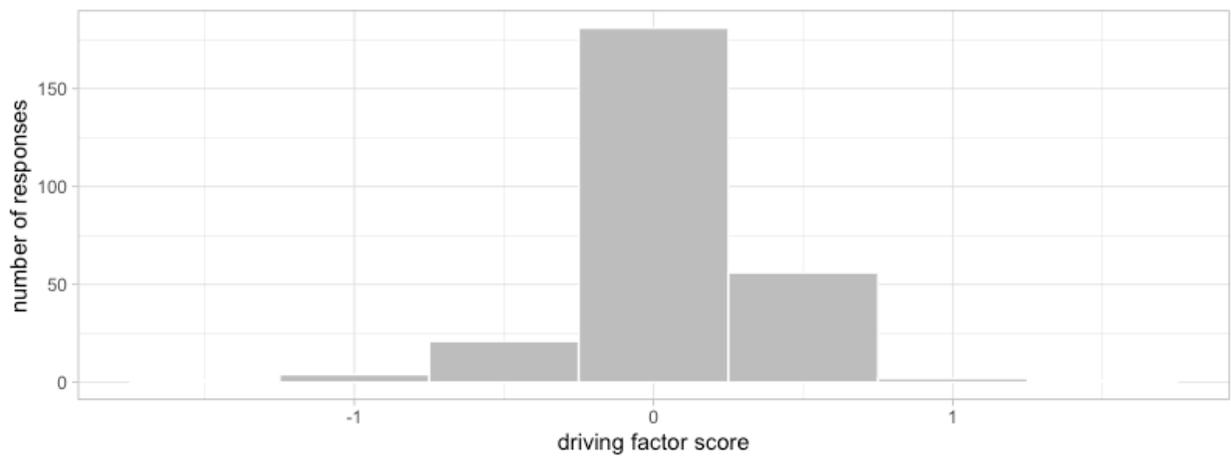
b. Driving

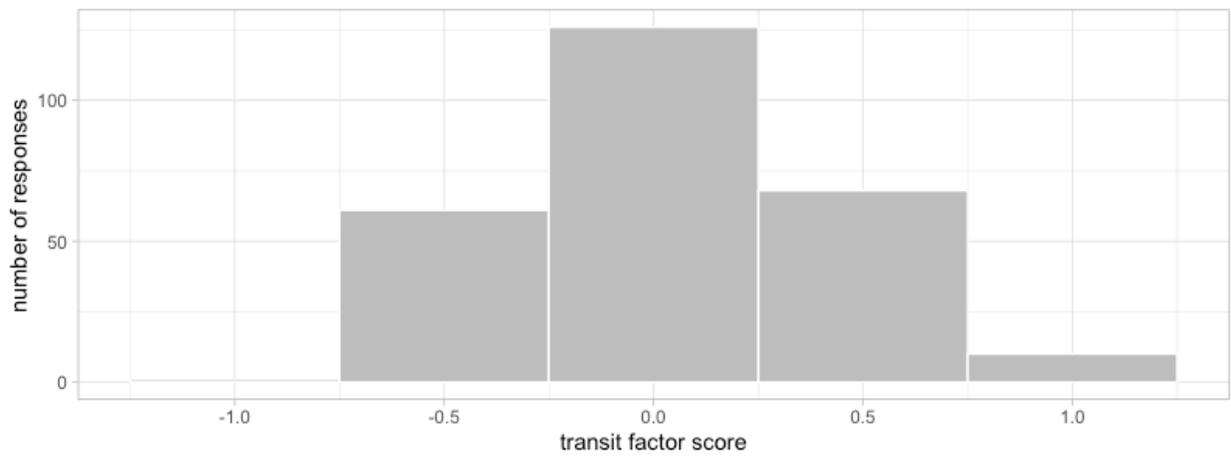
c. Transit

**Figure 4  Distribution of factor scores for three attitudinal aspects**

"normalized", indicating that owning a vehicle with driving assistance technologies and being happy about them might lead to a higher sentiment score. The "ride_bus" variable is positively correlated with "normalized", indicating that respondents who use transit are more positive about AVs and transit than those who do not use transit at all. Demographic variables such as "gen_m" and "ychildren" are positively correlated with "normalized", while "hh_size" is negatively correlated with "normalized." An ordered logistic regression was used to model the simplified sentiment score to estimate the respondents' opinions' relationship with other variables. The ordered logistic model fit was modest but consider sufficient to provide useful insights.

**Table 4  Descriptive statistics**

| Variable | Description | n | Mean | SD | Min | Max |
|---|---|---|---|---|---|---|
| sum | Sum sentiment score of each comment | 266 | 0.11 | 1.92 | -11 | 8 |
| normalized | Normalized sentiment score | 266 | 0.03 | 0.37 | -1 | 3 |
| simplified | Simplified sentiment score | 266 | 0.08 | 0.61 | -1 | 1 |
| av_bus_nd | OK to take AV bus w/o driver (no=-1, unsure=0, yes=1) | 265 | -0.23 | 0.82 | -1 | 1 |
| tech | Attitude toward technology (factor score) | 266 | -0.12 | 0.76 | -1.66 | 1.66 |
| driving | Attitude toward driving (factor score) | 266 | 0.06 | 0.30 | -1.26 | 1.26 |
| transit | Attitude toward transit (factor score) | 266 | 0.06 | 0.35 | -0.78 | 0.96 |
| datech_happy | Own and happy with driving assistance tech (no=0, yes=1) | 266 | 3.39 | 1.72 | 0 | 5 |
| ride_bus | Ride bus > once a week (no=0, yes=1) | 266 | 0.25 | 0.43 | 0 | 1 |
| veh_own | Household vehicle ownership (integer) | 266 | 1.81 | 0.96 | 0 | 4 |
| age | Age group (integer) | 266 | 4.58 | 1.59 | 1 | 8 |
| gen_m | Gender is male (no=0, yes=1) | 266 | 0.35 | 0.48 | 0 | 1 |
| education | Educational level (integer) | 263 | 4.68 | 1.24 | 1 | 6 |
| income | Income level (integer) | 227 | 3.35 | 1.50 | 1 | 6 |
| hh_size | Household size (number of people, integer) | 263 | 2.44 | 1.34 | 1 | 8 |
| ychildren | Number of kids (< 12 years old) in household (integer) | 260 | 0.47 | 1.19 | 0 | 11 |

Note: Age groups: 1: < 18; 2: 18-24; 3: 25-34; 4: 35-44; 5: 45-54; 6: 55-64; 7: 65-74; and 8: ≥ 75
   Educational levels: 1: Less than high school degree; 2: High school degree or equivalent;
      3: Some college but no degree; 4: Associate degree; 5: Bachelor's degree; and
      6: Graduate degree.
   Income levels: 1: < $25,000; 2: $25,000 - $49,999; 3 $50,000 - $74,999; 4: $75,000 - $99,999;
      5: $100,000 - $149,999"; and 6: ≥ $150,000.

The ordered logistic regression models are summarized in Table 6. One model is for the simplified sentiment and the other is for the willingness to take an automated bus without an operator. Both of these response variables have a "-1,0,1" scale, with -1 indicating negative sentiment or "not OK" answer for the willingness question, 0 indicating neutral sentiment or "neither OK nor not OK/unsure" about the willingness, and 1 indicating positive sentiment or "OK" about the willingness. The same set of explanatory variables as in the linear models were used in the logistic models. Model fit was examined using Lipsitz test, and the test statistic shows that the model fit was good. In the simplified sentiment model, variables "tech" and "income" are statistically significant at the 0.05 level; "ride_bus" and "age" are significant at the 0.1 level; "driving", "datech_happy", "gen_m", and "hh_size" are significant at the 0.15 level. In the automated bus model, variables "tech," "driving," "gen_m," and "education" are statistically significant at the 0.05 level.

**Table 5  Linear regression model of normalized sentiment score**

| Variable | Estimate | SE | p-value |
|---|---|---|---|
| (Intercept) | -0.068 | 0.158 | 0.668 |
| **tech** | -0.047 | 0.036 | **0.188** |
| **driving** | 0.158 | 0.094 | **0.095** |
| transit | 0.029 | 0.082 | 0.728 |
| **datech_happy** | 0.026 | 0.017 | **0.129** |
| **ride_bus** | 0.127 | 0.068 | **0.062** |
| veh_own | -0.037 | 0.037 | 0.313 |
| age | -0.001 | 0.017 | 0.932 |
| **gen_m** | 0.073 | 0.057 | **0.202** |
| education | -0.002 | 0.025 | 0.940 |
| income | 0.028 | 0.024 | 0.239 |
| **hh_size** | -0.036 | 0.026 | **0.172** |
| **ychildren** | 0.037 | 0.025 | **0.141** |

**Model fit** Multiple R-squared: 0.081; Adjusted R-squared: 0.029; F-statistic: 1.56 on 12 and 211 df, p-value: 0.105.

Note: SE = standard error; df = degree of freedom.
Variables with boldface are significant at the 0.2 level at least.

**Table 6  Ordered logistic regression models**

| Variable | Simplified sentiment model | | | Automated bus model | | |
|---|---|---|---|---|---|---|
| | Estimate | SE | p-value | Estimate | SE | p-value |
| **tech** | -0.454 | 0.186 | **0.014** | 0.676 | 0.190 | **0.000** |
| **driving** | 0.786 | 0.497 | **0.114** | 1.718 | 0.539 | **0.001** |
| transit | 0.102 | 0.420 | 0.809 | -0.530 | 0.435 | 0.223 |
| **datech_happy** | 0.133 | 0.088 | **0.129** | 0.074 | 0.087 | 0.392 |
| **ride_bus** | 0.593 | 0.348 | **0.088** | -0.133 | 0.342 | 0.697 |
| veh_own | -0.120 | 0.188 | 0.523 | -0.157 | 0.195 | 0.420 |
| **age** | -0.155 | 0.089 | **0.080** | -0.108 | 0.089 | 0.229 |
| **gen_m** | 0.439 | 0.293 | **0.135** | 1.172 | 0.296 | **0.000** |
| education | -0.090 | 0.126 | 0.477 | 0.299 | 0.132 | **0.023** |
| **income** | 0.257 | 0.122 | **0.035** | 0.004 | 0.122 | 0.974 |
| **hh_size** | -0.205 | 0.133 | **0.123** | 0.124 | 0.141 | 0.378 |
| ychildren | 0.152 | 0.124 | 0.220 | -0.026 | 0.150 | 0.861 |
| -1|0 | -1.812 | 0.835 | 0.030 | 1.381 | 0.823 | 0.093 |
| 0|1 | 1.170 | 0.828 | 0.158 | 2.838 | 0.840 | 0.001 |
| **Model fit** | Lipsitz test – H₀: the model is perfectly fit. LR statistic = 7.132, df = 9, p-value = 0.623 | | | Lipsitz test – H₀: the model is perfectly fit. LR statistic = 9.009, df = 9, p-value = 0.437 | | |

Note: H₀ = null hypothesis; LR = likelihood ratio.
Variables with boldface are significant at the 0.15 level at least.

The effects of some explanatory variables on "simplified" and "av_bus_nd" are visualized using effect plots as shown in Figure 5 and Figure 6, respectively. Not all explanatory variables were illustrated, only the ones that were of more interest were plotted. On a plot, the x-axis shows the value or level of an exploratory variable of interest, and the y-axis shows the probability of getting a certain value of the response variable.

In terms of simplified sentiment, there is a higher probability of neutral sentiment than negative or positive sentiment. The probability of neutral sentiment stays around 0.6, which is consistent with the survey data that over 60% of respondents were unsure about the future of AV-transit integration. Respondents that are more interested in technology are more likely to have more negative sentiments toward AV-transit integration than people that are less interested in technology. Respondents who like driving more tend to view AV-transit integration more positively than those who do not like driving as much. These results may be counterintuitive at first sight, but it is understandable that people who are more tech-savvy would have more concerns about new technologies, and people who love driving and interested in automobiles show more interests in AVs. Younger respondents tend to be more optimistic about AV-transit integration than older respondents. Respondents from households with higher combined income are more likely to have positively attitude toward AV-transit integration than those from lower-income households.

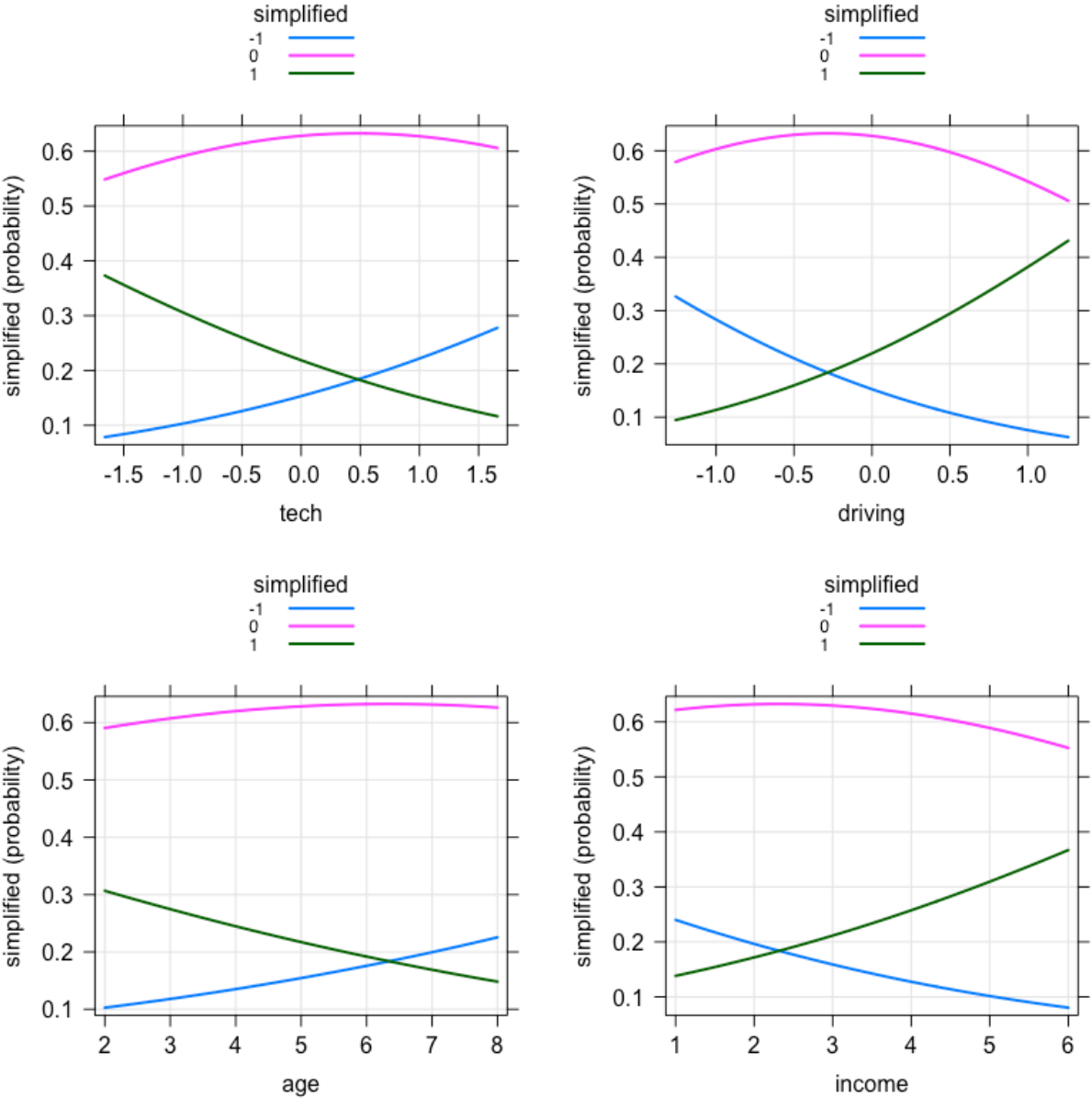

**Figure 5  Effect plots of simplified sentiment model**

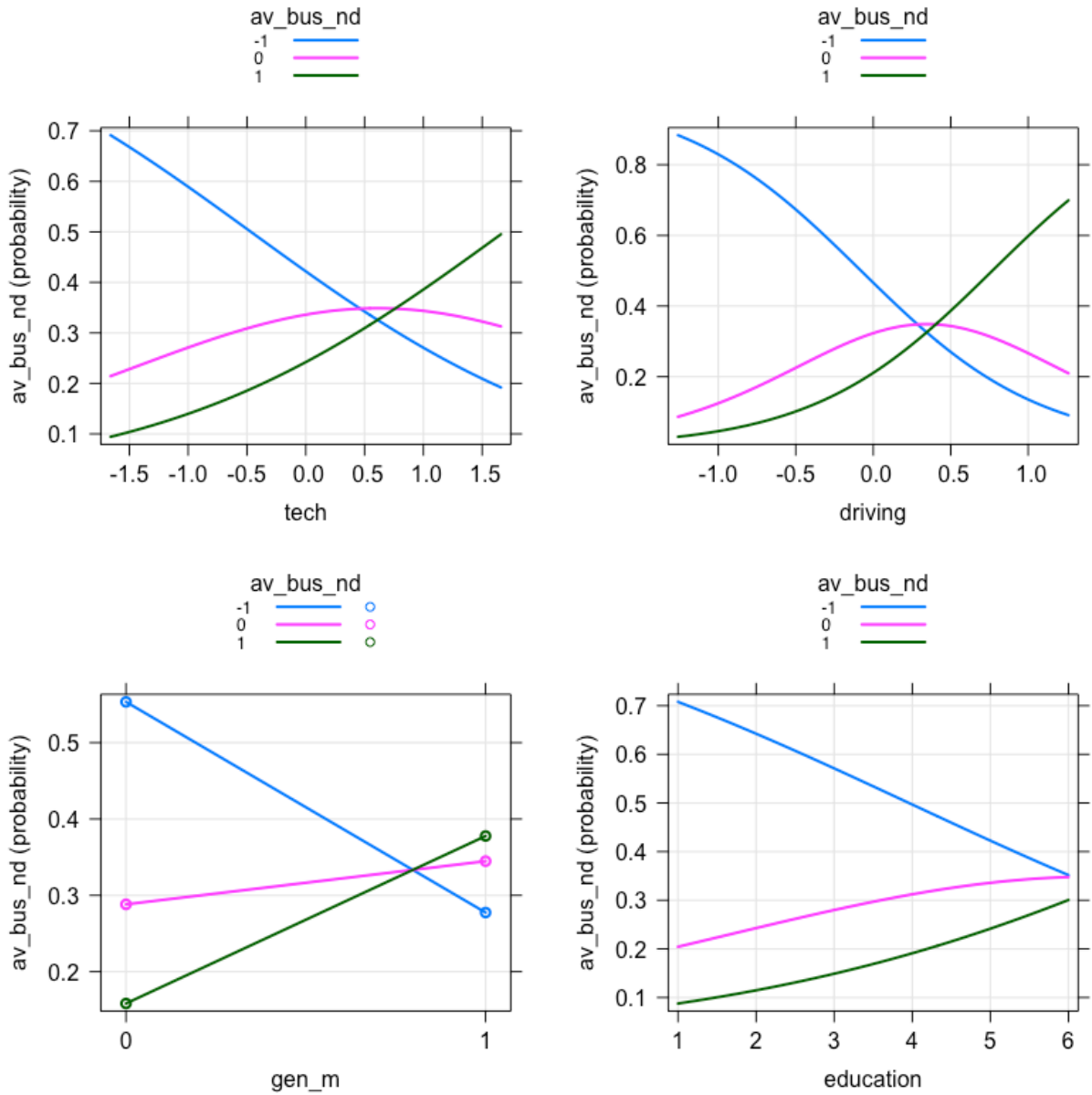

**Figure 6  Effect plots of automated bus model**

In terms of other significant variables, current driving assistance technology owners who are happy about those technologies or current transit users have more positive sentiments than those who do not own driving assistance technologies or those who do not use transit. Male respondents tend to have more positive sentiment toward AV-transit integration than those from other gender groups. Those who live in larger households are more positive toward AV-transit integration than respondents from smaller households.

Regarding the willingness to take an automated bus without a human operator onboard, the effect plots show that people that are more interested in technology tend to be more willing to ride an automated bus without a human operator. Respondents who like driving more are more willing to ride such a bus than people who are not as interested in driving. Male respondents are more willing to ride such a bus than those in other gender groups. Respondents with higher educational levels tend to be more willing to ride such a bus than those with lower educational levels.

## Conclusions

In this study, we analyzed data collected through an online survey carried out in two small urban areas in Wisconsin, to understand people's attitudes toward AVs and transit. Text mining, factor analysis, and regression analysis were used to obtain insights from 266 survey responses which consist of text, numerical, and categorical data. The analysis shows that a majority of small urban area residents are well exposed to some vehicle automation technologies and some information about AVs, but are unsure about what future vehicle automation and its integration with transit will look like. Major findings from this study are:

- Technology-savviness is a significant factor affecting people's attitudes toward AV-transit integration. Technology-savviness, among small urban area residents, is leading to excitement toward countless possibilities and improvements to transit that may be brought by vehicle automation. At the same time, technology savviness is also leading to concerns about problems that may be brought by vehicle automation.
- People's interest in driving leads to their interest in vehicle automation technologies and more positive opinions about AV-transit integration than people not as interested in driving.
- No significant relationship was found between people's attitudes toward transit and their opinions about AV-transit integration. However, current transit users did have significantly more positive opinions about AV-transit integration than those who do not use transit.
- There are differences in attitudes toward AV-transit integration among different groups of people with different demographics.

### *Comparison with Previous Studies*

For comparison, here are some key findings of the same aspects from previous studies on people's attitudes toward AVs and shared AVs. Previous studies generally suggested that a higher technology-savviness leads to a higher tendency to embrace AV technologies and use shared AVs (Bansal, Kockelman, and Singh 2016; Haboucha, Ishaq, and Shiftan 2017; Lavieri et al. 2017; Krueger, Rashidi, and Rose 2016). Results from this study is consistent in technology-savviness' leading to people's higher willingness to try AV technologies. However, we also found that technology-savvy people tend to have more concerns about potential problems brought by AVs and their disruption to transit.

Previous studies drew mixed conclusions about how people's interest in driving affects their opinions about AVs. Bansal, Kockelman, and Singh (2016) found that those who drive more were more likely to adopt AVs, and that they were more interested in owning Level-4 AVs than adding Level-3 automation or using shared AVs. Lavieri et al. (2017) concluded that individuals who currently own vehicles and who have not yet experienced carsharing services are more interested in adopting AVs. However, Haboucha, Ishaq, and Shiftan (2017) found that the respondents who enjoy driving are more likely to use their regular car than an AV. In this study, we found that in small urban areas, people who love driving have more positive attitudes toward AV-transit integration and are more willing to try automated transit vehicles than those who are not as interested in driving.

In terms of how people's attitudes toward transit affecting their opinion and use of AVs, previous studies mostly agreed that transit users or people who have positive attitudes toward transit are more likely to use shared AVs (Haboucha, Ishaq, and Shiftan 2017; Lavieri et al. 2017).

However, transit users still have concerns about the operational and personal safety for riding an automated bus (Dong, DiScenna, and Guerra 2019). In this study, although no significant relationship was found between people's attitudes toward transit and their opinions about AV-transit integration, current transit users showed significantly more positive opinions about AV-transit integration than non-transit users. Therefore, this finding for small urban areas is consistent with the findings for large urban areas from previous studies.

Differences in travel behaviors may help explain some of the differences in attitudes toward AVs and transit between people in small urban areas and people in large urban areas. Compared with residents in large urban areas, travels made by small urban area residents are more "car dependent". Over 90% of the surveyed small urban area residents use personal vehicle as their major travel mode. However, people in small urban areas, although are more car-dependent, generally do not hold negative views toward transit as people in large urban areas. A clear boundary may exist between "transit supporters" and "car lovers" in large urban areas, but there is no such a clear boundary in small urban areas. The survey results showed that people love driving more are even more willing to ride an automated bus than those less interested in driving. In the studied small urban areas, the top reasons that prevented people from using transit were lack of convenience, lack of flexibility, and poor access. Also, current transit users in these small urban areas do have a more positive opinion about future AV-transit integration. Therefore, once the transit service is improved with the help of AVs, it will attract more people to use it.

*Limitations and Recommendations*

Limitations of this study should be noted, and directions for future work need to be pointed out. The findings from this study are based on data collected in two small urban areas in Wisconsin. A broader data collection from a wide spectrum of small urban areas across the nation is necessary to validate these findings. This study is novel in its combinatory use of analytical methods including text mining, factor analysis, and regression analysis. We managed to make the most use of those analytical methods to get very useful insights from the survey data, but due to the limitations in sample size and questionnaire design, some insights may not be supported by perfectly fit models or statistically significant results. However, we believe this study serves well as a foundation for future expanded attitudinal survey study on the topic of AV-transit integration.

Small urban areas will be an important market for AVs. Small urban areas make 2/3 of all urban areas in the United States. Also, there exists a gap between the demand for better transit services and the under-supply of transit services, which can be filled by AVs. There is a great potential in integrating AVs with transit services in small areas. AVs, if well-integrated with transit services, will be able to satisfy the flexible travel needs of these small urban area residents. Such an integration may also help change these residents' travel behavior and reduce the overall car ownership.

## Acknowledgements
The authors thank the Tommy G. Thompson Center on Public Leadership for sponsoring this study, and the cities of Eau Claire and Janesville for collaborating on the survey.